\documentclass[10pt,twocolumn,letterpaper]{article}
\usepackage{wacv}  
\usepackage{graphicx}
\usepackage{bm} 
\usepackage{amsmath}
\usepackage{amssymb}
\usepackage{booktabs}
\usepackage{makecell}  % for \makecell command
\usepackage{multirow}
\usepackage{textcomp} % for \textsuperscript command
\usepackage{comment}
\usepackage[utf8]{inputenc} % allow utf-8 input
\usepackage[T1]{fontenc}    % use 8-bit T1 fonts

\usepackage[pagebackref,breaklinks,colorlinks]{hyperref}

\usepackage[capitalize]{cleveref}
\crefname{section}{Sec.}{Secs.}
\Crefname{section}{Section}{Sections}
\Crefname{table}{Table}{Tables}
\crefname{table}{Tab.}{Tabs.}

\def\wacvPaperID{2097} % *** Enter the WACV Paper ID here
\def\confName{WACV}
\def\confYear{2025}

\DeclareMathOperator*{\argmin}{arg\,min}

\begin{document}

%%%%%%%%% TITLE - PLEASE UPDATE
\title{CUNSB-RFIE: Context-aware Unpaired Neural Schr\"{o}dinger Bridge in Retinal Fundus Image Enhancement}

\author{ Xuanzhao Dong$^{*}$\\
Arizona State University\\
{\tt\small xdong64@asu.edu} \\
\and
Vamsi Krishna Vasa$^{*}$\\
Arizona State University\\
{\tt\small vvasa1@asu.edu}\\
\and 
Wenhui Zhu\\
Arizona State University\\
{\tt\small wzhu59@asu.edu}\\
\and
Peijie Qiu \\
Washington University in St.Louis\\
{\tt\small peijie.qiu@wustl.edu}\\
\and 
Xiwen Chen\\
Clemson University\\
{\tt\small xiwenc@g.clemson.edu}
\and
Yi Su\\
Banner Alzheimer's Institute\\
{\tt\small Yi.Su@bannerhealth.com}
\and
Yujian Xiong\\
Arizona State University\\
{\tt\small yxiong42@asu.edu}\\
\and
Zhangsihao Yang\\
Arizona State University\\
{\tt\small zshyang1106@gmail.com}\\
\and 
Yanxi Chen\\
Arizona State University\\
{\tt\small ychen855@asu.edu}\\
\and 
Yalin Wang\\
Arizona State University\\
{\tt\small ylwang@asu.edu}
% For a paper whose authors are all at the same institution,
% omit the following lines up until the closing ``}''.
% Additional authors and addresses can be added with ``\and'',
% just like the second author.
% To save space, use either the email address or home page, not both
}
\maketitle

%%%%%%%%% ABSTRACT
\begin{abstract}
    \noindent Retinal fundus photography is significant in diagnosing and monitoring retinal diseases. However, systemic imperfections and operator/patient-related factors can hinder the acquisition of high-quality retinal images. Previous efforts in retinal image enhancement primarily relied on GANs, which are limited by the trade-off between training stability and output diversity. In contrast, the Schr\"{o}dinger Bridge (SB), offers a more stable solution by utilizing Optimal Transport (OT) theory to model a stochastic differential equation (SDE) between two arbitrary distributions. This allows SB to effectively transform low-quality retinal images into their high-quality counterparts. In this work, we leverage the SB framework to propose an image-to-image translation pipeline for retinal image enhancement. Additionally, previous methods often fail to capture fine structural details, such as blood vessels. To address this, we enhance our pipeline by introducing Dynamic Snake Convolution, whose tortuous receptive field can better preserve tubular structures. We name the resulting retinal fundus image enhancement framework the Context-aware Unpaired Neural Schr\"{o}dinger Bridge (CUNSB-RFIE). To the best of our knowledge, this is the first endeavor to use the SB approach for retinal image enhancement. Experimental results on a large-scale dataset demonstrate the advantage of the proposed method compared to several state-of-the-art supervised and unsupervised methods in terms of image quality and performance on downstream tasks.The code is available at \url{https://github.com/Retinal-Research/CUNSB-RFIE}.
\end{abstract}

\def\thefootnote{*}\footnotetext{These authors contributed equally to this paper.}

\begin{figure}[htbp]
  \centering
  \includegraphics[width=1\columnwidth]{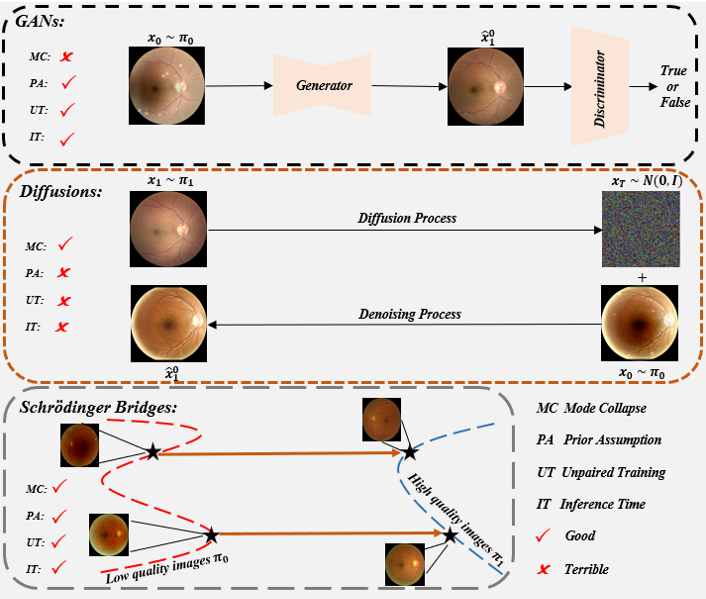}
  \caption{Three generative model pipelines in retinal image enhancement: \textbf{GANs} face the challenge of balancing training stability with generation diversity and quality. \textbf{Diffusion models} require paired data from the source domain as a condition and are limited by the prior Gaussian assumption and longer inference times. \textbf{SB} does not suffer from these limitations.}
  \label{CUNSB-intuitive}
\end{figure}

%%%%%%%%% BODY TEXT
\section{Introduction}
\label{sec:intro}

\noindent Retinal color fundus photography (CFP) is crucial for diagnosing ocular diseases \cite{deeplearning1,deeplearning2,deeplearning3,deeplearning4,deeplearning5}. However, various human-related or equipment-related factors (e.g., variability in surrounding light conditions, image capture errors, lens contamination) introduce complex (typically multimodal) noise, compromising image quality and obscuring vital clinical details such as blood vessels and lesions. These factors limit the use of conventional physical models \cite{41493}, which are often designed to address specific types of degradation, such as those caused by internal cataractous media. Therefore, it is indispensable to develop a unified, automated, and robust technique to improve CFP quality. This sequentially enhances retinal disease diagnosis (e.g., diabetic retinopathy) and develops better automatic screening tools. 

\noindent Recently, deep learning-based methods have achieved state-of-the-art performance.  However, due to the limited availability of paired clean-noisy retinal images, which are difficult and costly to collect in practice, the focus has shifted from supervised methods~\cite {shen2020modeling,10.1007/978-3-031-16434-7_49,li2023generic} to unsupervised methods. In particular, Generative Adversarial Networks (GANs) \cite{goodfellow2014generative} have been widely utilized in previous studies for fundus image enhancement, formulating it as an image-to-image (I2I) translation problem~\cite{9763342,zhu2023optimal,zhu2023otre}. These models typically learn a direct translation mapping using an adversarial learning strategy. However, this approach tends to be suboptimal (e.g., mode collapse) when dealing with images that involve complex structures or exhibit multimodal distributions~\cite{salmona2022can}. Denoising diffusion probabilistic models (DDPMs)~\cite{ho2020denoising,sohl2015deep,song2020denoising} present a promising alternative, as they are capable of learning complex distribution and generating diverse, high-quality images~\cite{xiao2021tackling} by leveraging their iterative nature. However, DDPMs face limitations due to their reliance on Gaussian noise prior assumptions~\cite{ho2020denoising} and longer inference times~\cite{selim2023latent,shao2024prior}. Additionally, by modeling them as a single domain, their requirement for paired conditional data limits their applicability in image-to-image translation tasks~\cite{rombach2022high}.

\noindent Schr\"{o}dinger Bridges (SBs) address the limitation of prior distributions in DDPMs while preserving the smooth and probabilistically consistent transformation between distributions. SBs achieve this finding through a stochastic process between two arbitrary distributions by solving an entropy-regularized optimal transport (OT) problem. Their flexibility in prior distributions has motivated a wide spectrum of applications in medical imaging~\cite{li2024diffusion,wang2024implicit}. However, both methods require paired data to learn stochastic differential equations (SDEs). To the best of our knowledge, none of these works has successfully trained SBs to solve the retinal fundus enhancement task in an unpaired fashion.

\noindent Another major challenge in retinal image enhancement is preserving crucial structural information, such as blood vessels, between low-quality images and their high-quality counterparts. Although score-based diffusion models excel in generating natural images, they may lose track of structural representation during the iterative process due to the uniform injection of Gaussian noise to every pixel in retinal images. The Dynamic Snake Convolution (DSC)~\cite{qi2023dynamic}, which modifies standard square filters into curving structures, can be a promising compensation for the model drawbacks. These 'snake-shaped' convolutions help the generator focus more on tiny vessel structures, preventing them from being overlooked as the state evolves. Additionally, to prevent over-tampering with important structures and to preserve the consistency between low-quality and enhanced images, we apply PatchNCE~\cite{park2020contrastive} and SSIM~\cite{brunet2011mathematical} regularization during the training process. PatchNCE regularization encourages patches in the target domain to be generated from their corresponding patches in the source domain within the feature space. In contrast, SSIM regularization promotes structure similarity at the image level by preserving local patterns and textures.   

\noindent Our main contributions are threefold: \textbf{(i)} We propose a novel SBs-based unpaired retinal image enhancement paradigm that retains the benefits of the iterative nature in DDPMs while eliminating their prior distribution assumptions and long inference steps.
\textbf{(ii)} We introduce CUNSB-RFIE, which leverages Dynamic Snake Convolution, PatchNCE and SSIM regularizations to preserve contextual information in retinal images.
\textbf{(iii)} We conduct comprehensive experiments on three large publicly available retinal fundus datasets and demonstrate the goodness of our methods over multiple supervised and unsupervised strategies.

\section{Related Work}
\noindent \textbf{Fundus Image Enhancement.} Early state-of-the-art Neural-based exploration to fundus image enhancement leveraged supervised and self-supervised methods. Shen et al. \cite{shen2020modeling} introduced Cofe-Net, which directly transferred noisy images to their clean versions with the help of related segmentation map. Liu et al. \cite{10.1007/978-3-031-16434-7_49} introduced PCE-Net, which uses Laplacian pyramid constraints decomposed from low-quality images to guide the enhancement process. To further reduce the need for paired data, GFE-Net~\cite{li2023generic} and SCR-Net~\cite{scrnet} were proposed, enhancing images by using frequency information and synthetic data. However, paired data was still required during their training process, which is unwished in practice.

\noindent To fully overcome this limitation, unsupervised methods such as GANs~\cite{goodfellow2014generative} have been widely adopted. Specifically, CycleGAN~\cite{cyclegan} serves as the most common enhancement strategy by modeling the task as an image-to-image (I2I) translation problem. Its unpaired training pipeline and cycle-consistent regularization eliminate the need for paired data and preserve the structure consistency, respectively, albeit with the introduction of additional computational cost and a large search space of parameters. Building on this, Li et al.~\cite{li2022annotation} proposed ArcNet, which reduces the search space by focusing on high-frequency components. Zhu et al.~\cite{zhu2023optimal,zhu2023otre} proposed using the Structural Similarity Index (SSIM) constraint to prevent over-tampering of vessel and lesion structure in fundus images. Despite the impressive performance demonstrated by previous GAN-based enhancement strategies, they typically require additional techniques for training stability, such as gradient penalty and spectral normalization, to ensure the Lipschitz continuity of the discriminator. While these techniques stabilize training, they may restrict the network's ability to learn complex distributions, potentially contributing to suboptimal issues like mode collapse~\cite{salmona2022can}.

\noindent Diffusion models, which typically involve contaminating the images by sequentially adding Gaussian noise and then generating images through a backward denoising process, tend to be an alternative choice due to their ability to learn multimodel distributions of images, leading to diverse and high-quality results~\cite{xiao2021tackling,salmona2022can}. Its strong ability in image generation has motivated work on retinal image enhancement. For instance, Liu \& Huang~\cite{liu2023esdiff} proposed a multi-task diffusion pipeline to simultaneously address quality enhancement and vessel segmentation problems. However, the slow inference speed and Gaussian prior assumption remain significant concerns. Additionally, using the source domain image as the condition information for performing I2I tasks presents another drawback, since it usually requires paired input-condition data, which is often infeasible in practice. Although some works~\cite{gu2023optimal,wu2023latent} have attempted to address this issue, their performance on retinal image enhancement remains unclear. Schr\"{o}dinger Bridges (SBs) thus present a promising third solution, as they are free from prior distribution assumption and allow for fast inference. To the best of our knowledge, our work is the first to use SBs to directly enhance retinal fundus images.

\noindent \textbf{Schr\"{o}dinger Bridges}. The Schr\"{o}dinger Bridge (SB) 
problem~\cite{schrodinger1932theorie,leonard2013survey} aims to find the stochastic process that guides a system from an initial distribution to a target distribution over time, subject to a reference measure. Attracted by its flexibility in selecting boundary distributions, many works have theoretically and practically focused on SBs. In terms of algorithms, Bortoli et al.~\cite{de2021diffusion} unified its relationship with score-based diffusion by leveraging Iterative Proportional Fitting (IPF) algorithms and Vargas et al.~\cite{vargas2021solving} proposed to approximate SBs using Gaussian Processes. 
In applications, Liu et al.~\cite{liu20232} proposed $I^2SB$, which solves multiple vision tasks using clean-degraded pairs to learn SBs.
Shi et al.~\cite{shi2024diffusion} went beyond paired scenarios in image transformation, learning SBs leveraged by Markovian projection and Markov measures. Kim et al.~\cite{kim2023unpaired} were the first to decompose SB problems into a sequence of adversarial learning processes to solve high-resolution I2I tasks, which inspired our work. This paper shows the potential of SB-based models in medical I2I tasks and paves the way for their future usage in various applications.

%-------------------------------------------------------------------------
\begin{figure*}[!t]
  \centering
  \includegraphics[width=0.9\textwidth]{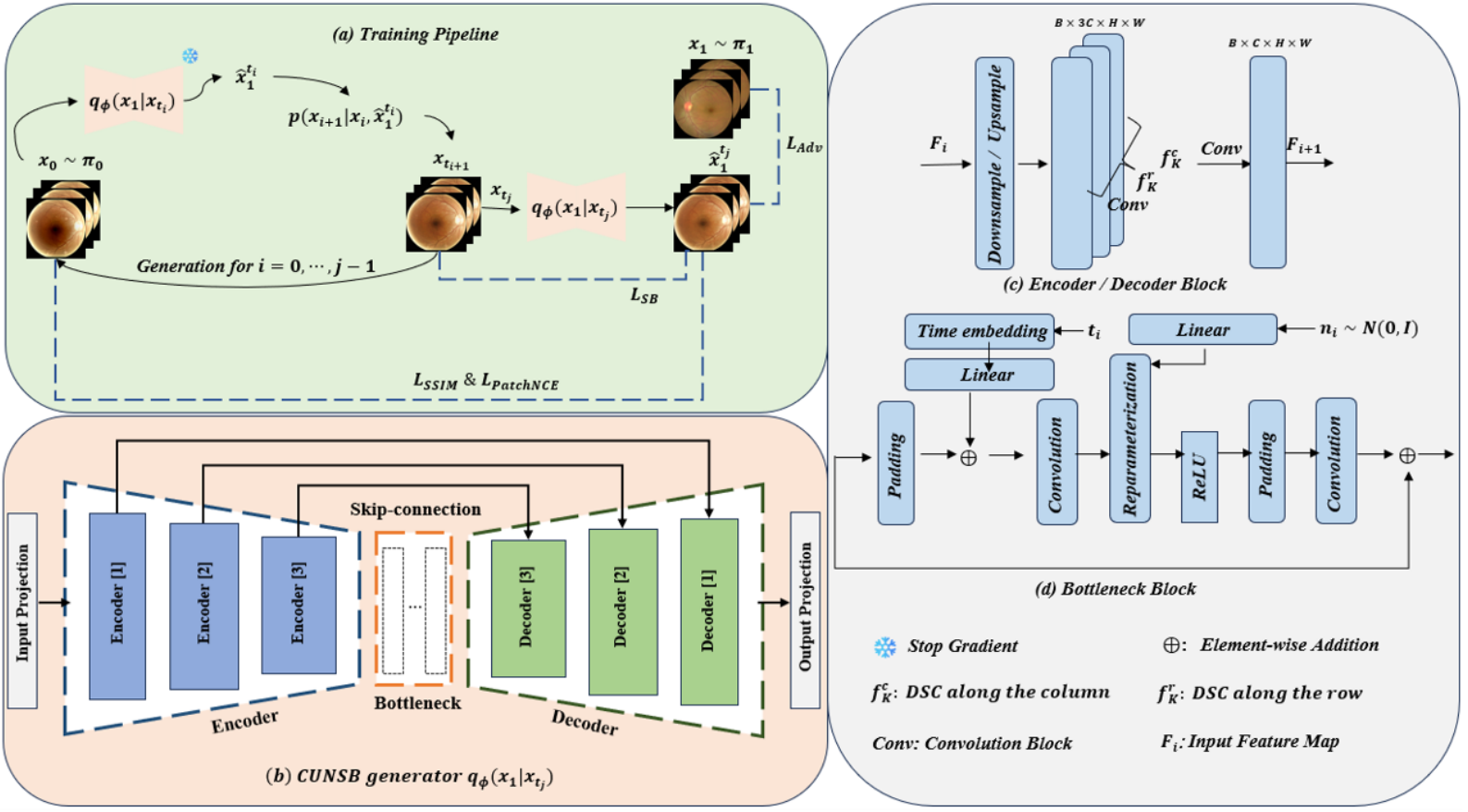}  % Replace with your figure file and remove the example
  \vspace{-0.4cm}
  \caption{Illustration of our CUNSB-RFIE framework. (\textbf{a}) The training pipeline. (\textbf{b}) The U-Net-like generator structure. (\textbf{c}) The Encoder and Decoder block structure in the generator with Dynamic Snake Convolution. (\textbf{d}) The structure of the Bottleneck block with Gaussian noise and time information embedding.} 
  \label{CUNSB-structure}
\end{figure*}

\section{Method}

\subsection{Background}\label{sec:back}
\noindent \textbf{Schr\"{o}dinger Bridges}. Given two probability distribution  $\pi_0$, $\pi_1$ on state space $\mathbb{R}^d$, the Schr\"{o}dinger Bridges Problem (SBP) seeks to find the optimal stochastic process $\{\bm{x}_t: t \in [0,1]\}$ that evolves from initial distribution $\pi_0$ at $t = 0$ to terminal distribution $\pi_1$ at $t = 1$. Specifically, let $\Omega$ denote the path space on $\mathbb{R}^d$, representing all possible continuous functions mapping the time interval $t \in [0, 1]$ to the state space, and let $\mathbb{Q} \in \mathcal{P}(\Omega)$ represent a probability measure on the defined path space. The SBP aims to solve the following optimization problem subject to certain constraints:

\begin{equation} 
\begin{split}
\mathbb{Q}^{SB} &= \argmin_{\substack{\mathbb{Q} \in \mathcal{P}(\Omega)}} \mathbb{D}_\mathrm{KL}(\mathbb{Q} \| \mathbb{W}) \\
 &\text{subject to} \ \mathbb{Q}_0 = \pi_0, \ \mathbb{Q}_1 = \pi_1  \label{eq:sbp}
\end{split}
\end{equation}

\noindent where $\mathbb{W}$ represents the reference measure, typically chosen to be the Wiener measure with variance, denoted by $\mathbb{W}^\tau$, and $\mathbb{Q}_t$ represents the marginal distribution of $\mathbb{Q}$ at time t. The optimal measure of Eq.~\ref{eq:sbp}, denoted by $\mathbb{Q}^{SB}$, is called the Schrödinger Bridge (SB) between $\pi_0$ and $\pi_1$. As an unpaired image-to-image translation task, the retinal fundus image enhancement mission can be modeled as finding the optimal evolution of states from the source low-quality image domain $x_0 \sim \pi_0$  to the high-quality image domain $x_1\sim \pi_1$, where $\pi_0$ and $\pi_1$ represent the boundary image distributions.

\noindent \textbf{Static Formulation and Self-similarity}. Since the possible measures are indexed by the time step t, tracking the optimal $\mathbb{Q}$ throughout the entire process can be challenging. The static formulation of the SB provides a way to simplify the problem to a more tractable form. Specifically, let $\Pi(\pi_0,\pi_1)$ represent a set of all possible joint distributions with marginals at $t=0$ and $t=1$ being $\pi_0$ and $\pi_1$, respectively, and let $H( \cdot )$ represents the entropy function. The Schr\"{o}dinger Bridges Coupling between the boundary distributions, denoted by $\mathbb{Q}^{SB}_{01}$, can be viewed as the solution to an entropy-regularized optimal transport problem, leading to the same marginal distribution of $\mathbb{Q}^{SB}$~\cite{tong2023improving} such that:
\begin{equation}
    \begin{split}
        \mathbb{Q}^{SB}_{01} = \argmin_{\gamma \in \Pi(\pi_0,\pi_1)} \mathbb{E}_{(\bm{x}_0,\bm{x}_1) \sim \gamma} [\|\bm{x}_0 - \bm{x}_1\|^2] - 2 \tau H(\gamma) \label{eq:static_formu}
    \end{split}
\end{equation}
\noindent Furthermore, the marginal distribution of $\bm{x}_t$ conditioned on $\bm{x}_0,\bm{x}_1$ follows a Gaussian distribution:
\begin{equation}
    \begin{split}
        p(\bm{x}_t | \bm{x}_0,\bm{x}_1) \sim \mathcal{N}(\bm{x}_t ; t \bm{x}_1 + (1 - t) \bm{x}_0, t(1 - t) \tau \bm{I}) \label{eq: sb_dis}
    \end{split}
\end{equation}
\noindent where $\{\bm{x}_t\} \sim \mathbb{Q}^{SB}, t \in [0, 1]$ and $\bm{x}_0 \sim \pi_0, \bm{x}_1 \sim \pi_1$. The concept of self-similarity~\cite{kim2023unpaired} extends the Eq. \ref{eq:static_formu} and Eq.~\ref{eq: sb_dis} to arbitrary sub-intervals. Specifically, for $[t_a,t_b] \subseteq [0,1]$ and $\{\bm{x}_t\} \sim \mathbb{Q}^{SB}$, the Schr\"{o}dinger Bridges Coupling between $t_a$ and $t_b$, denoted by $\mathbb{Q}^{SB}_{t_at_b}$, can be obtained as follows:
\begin{align}
        \mathbb{Q}^{SB}_{t_at_b} &= \argmin_{\gamma \in \Pi(\mathbb{Q}_{t_a},\mathbb{Q}_{t_b})} \mathbb{E}_{(\bm{x}_{t_a},\bm{x}_{t_b}) \sim \gamma} [\|\bm{x}_{t_a} - \bm{x}_{t_b}\|^2] \nonumber \\
        & - 2 \tau (t_b-t_a) H(\gamma) \label{eq:static_formu_any}
\end{align}
\noindent Then the marginal distribution of $\bm{x}_t$ for $t \in [t_a,t_b]$ conditioned on $ \bm{x}_{t_a}$ and $ \bm{x}_{t_b}$ still follows a Gaussian distribution, such that:
\begin{equation}
    \begin{split}
        p(\bm{x}_t | \bm{x}_{t_a},\bm{x}_{t_b}) \sim \mathcal{N}(\bm{x}_t ; \mu_{t_a,t_b}, \sigma_{t_a,t_b}) \label{eq: sb_dis_any} \\
        \text{s.t.} \begin{cases}
            \mu_{t_a,t_b} = s(t) \bm{x}_{t_b} + (1 - s(t)) \bm{x}_{t_a} \\
            \sigma_{t_a,t_b} =  s(t)(1 - s(t)) \tau (t_b -t_a) \bm{I} \\
        \end{cases} 
    \end{split}
\end{equation}
\noindent where $s(t):= (t-t_a)/(t_b-t_a)$. Eq.~\ref{eq: sb_dis_any} and Eq.~\ref{eq: sb_dis} provide effective methods to sample $\{\bm{x}_t\}$ from $\mathbb{Q}^{SB}$ for simulation.

\subsection{Context-aware Unpaired Neural Schr\"{o}dinger Bridge}\label{sec:drsb}
\noindent \textbf{Schr\"{o}dinger Bridge Learning}. To learn the Schr\"{o}dinger Bridge (SB) between low-quality and high-quality image domains, we implement the concepts from~\cite{kim2023unpaired}. Specifically, let $q_{\phi}(\bm{x_1}|\bm{x}_{t_i})$ be our 
 target generator, parameterized by $\phi$, which accepts the input $\bm{x}_{t_i}$ and $t_i$. $\{t_i\}_{i = 0}^N$ represents a given time partition of $[0,1]$ . Learning the SB then can be expressed as an optimization problem to approximate its static formulation shown in Eq.~\ref{eq:static_formu_any} over an arbitrary sub-interval $[t_i,1]$, such that
\begin{align} \label{eq:optim} 
    \min_{\phi} & \quad \mathbb{L}_{SB}(\phi,t_i) := \mathbb{E}_{q_{\phi}(\bm{x}_{t_i},\bm{x}_1)} [\|\bm{x}_{t_i} - \bm{x}_1\|^2] \nonumber  \\
    & \quad - 2 \tau (1 - t_i) H(q_{\phi}(\bm{x}_{t_i}, \bm{x}_1)) \nonumber \\
\text{s.t.} & \quad \mathbb{L}_{Adv}(\phi,t_i) := \mathbb{D}_\mathbb{KL}(q_{\phi}(\bm{x}_1) \| p(\bm{x}_1)) = 0 
\end{align}
\noindent where the constraint ensures that our generator actually learns the high-quality image distribution. By applying Lagrange Multiplier, Eq.~\ref{eq:optim} can be modified into the loss function $\mathbb{L}$ for a given $t_i$, such that:
\begin{align}
    \min_{\phi} \mathbb{L}(\phi,t_i) := \mathbb{L}_{Adv}(\phi,t_i) + \lambda_{SB} \mathbb{L}_{SB}(\phi,t_i). \label{eq:UNSB}
\end{align}
\noindent The relationship between $q_{\phi}(\bm{x_1},\bm{x}_{t_i})$ and $q_{\phi}(\bm{x_1}|\bm{x}_{t_i})$ is defined as:
\begin{equation}
    q_{\phi}(\bm{x_1},\bm{x}_{t_i}) := q_{\phi}(\bm{x_1}|\bm{x}_{t_i})p(\bm{x}_{t_i}).
\end{equation}
The true marginal distribution $p(\bm{x}_{t_i})$ can be approximated by its Markovian decomposition, where each component $p(\bm{x}_{t_{j+1}} | \bm{x}_{t_j}) = q_{\phi}(\bm{x}_{t_{j+1}} | \bm{x}_{t_j})$ for $j$ in $\{0,1,\cdots, i-1 \}$ ($p(\bm{x}_0)\sim \pi_0$) is approximated as:
\begin{equation}
    q_{\phi}(\bm{x}_{t_{j+1}} | \bm{x}_{t_j}) := \mathbb{E}_{q_{\phi}(\bm{x}_1 | \bm{x}_{t_j})}[p(\bm{x}_{t_{j+1}} | \bm{x}_{1}, \bm{x}_{t_j})]
\end{equation}
once Eq.~\ref{eq:UNSB} can be solved by $\phi$ and $p(\bm{x}_{t_{j+1}} | \bm{x}_{1}, \bm{x}_{t_j})$ can be approximated using Eq.~\ref{eq: sb_dis_any} after simply modifying the terminal condition such that $t_b =1, t_a=t_j$. Training and inference steps can be performed under the same framework as in~\cite{kim2023unpaired} with more details shown in Fig.~\ref{CUNSB-structure}.

\begin{figure}[htbp]
  \centering
  \includegraphics[width=0.8\columnwidth]{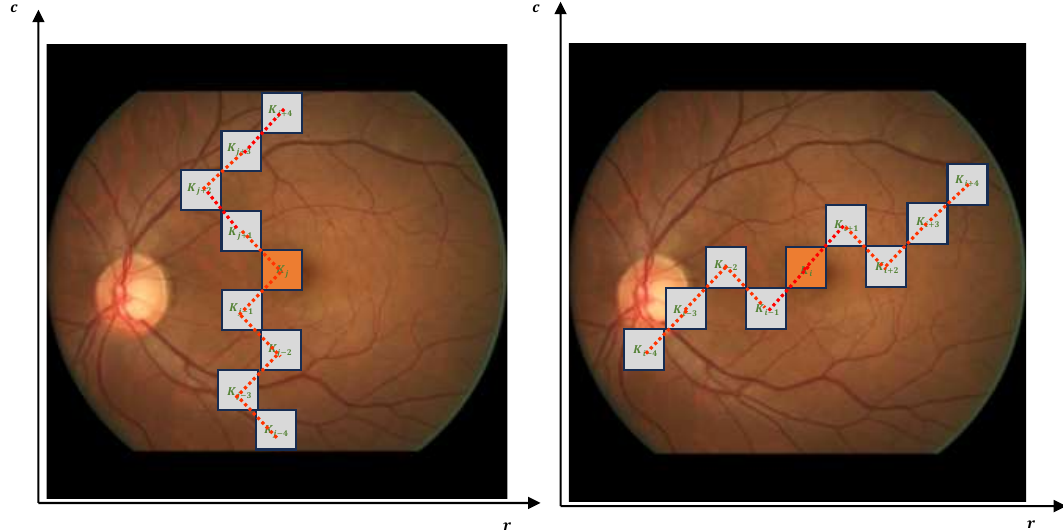}
  \caption{\textbf{Left}. The DSC coordinate grids along the column \textbf{Right}. The DSC coordinate grids along the row}
  \label{CUNSB-DSC}
\end{figure}

\noindent \textbf{Tubular Feature Capture}. Tubular structures, such as blood vessels, are abundant and crucial in retinal fundus images. However, our experiments show that the vanilla UNet model fails to retain contractual information in the synthetic high-quality counterparts, largely due to the conventional convolutional neural layer having difficulty capturing the vessel morphology. Recently, various works have been proposed to enhance the transformation modeling capability of convolutional neural networks (CNNs) by adjusting the spatial sampling locations in the deformable convolutional kernel~\cite{dai2017deformable,yu2017dilated,qi2023dynamic}, thus addressing these limitations. Inspired by these works, we introduced Dynamic Snake Convolution (DSC) into the generator, which helps the model to capture the vessel structures. 

\noindent Specifically, let $\mathbb{K}$ represent the straightened 2D convolution kernel of size 9, where the central coordinate is denoted by $\mathbb{K}_i=(\bm{r}_i,\bm{c}_i)$. The surrounding points in the grid are represented as $\mathbb{K}_{i \pm z} = (\bm{r}_{i \pm z},\bm{c}_{i \pm z})$ for $z \in \{0,1,2,3,4\}$. The DSC along the row considers the following coordinates in the kernel, denoted as: 
\begin{equation}\label{eq:xsnake}
\mathbb{K}_{i \pm z} =
    \begin{cases}
         (\bm{r}_{i+z},\bm{c}_{i+z}) = (\bm{r}_i+z,\bm{c}_i+\Sigma_i^{i+z}\Delta\bm{c})  \\
         (\bm{r}_{i-z},\bm{c}_{i-z}) = (\bm{r}_i-z,\bm{c}_i+\Sigma_{i-z}^{i}\Delta\bm{c}), \\
    \end{cases} 
\end{equation}
\noindent where $\Delta\bm{c} \in [-1,1]$ represents learnable deformation offsets along the column direction. The same operation is then applied along the column direction to the following coordinates, denoted as:
\begin{equation}\label{eq:ysnake}
\mathbb{K}_{j \pm z} =
    \begin{cases}
         (\bm{r}_{j+z},\bm{c}_{j+z}) = (\bm{r}_j+\Sigma_j^{j+z}\Delta\bm{r},\bm{c}_j+z)  \\
         (\bm{r}_{j-z},\bm{c}_{j-z}) = (\bm{r}_j+\Sigma_{j-z}^{j}\Delta\bm{r},\bm{c}_j-z),  \\
    \end{cases} 
\end{equation}
\noindent Since the offsets $\Delta\bm{c},\Delta\bm{r}$ result in fractional coordinates, bilinear interpolation is used to perform the convolution operation. Further details are illustrated in Fig.~\ref{CUNSB-DSC}. Let $\mathbb{F}_i$ represent the input feature map at each stage of the encoder and decoder blocks in the generator $q_\phi$. The DSC can then be formally expressed as:
\begin{equation}\label{eq:DSCstructure}
    \mathbb{F}_{i+1} = DSC(\mathbb{F}_i).
\end{equation}
The detailed placement strategy is depicted in Fig.~\ref{CUNSB-structure} (c). Here, DSC results in a curvilinear receptive field that adapts to the geometric structure of blood vessels in retinal images. Consequently, our generator will focus more on learning this semantic information during the training phase, leading to improved performance in preserving vessel structure.

\noindent \textbf{Context-preserving regularization}. Let $\hat{\bm{x}}_{1}^{t_i}$ represents samples from the generator $q_{\phi}(\bm{x_1}|\bm{x}_{t_i})$ at time $t_i$. The solution to Eq.~\ref{eq:UNSB} provides a straightforward method for generating synthetic high-quality images, denoted as  $ \hat{\bm{x}}_{1}^0 $. To ensure that the generator $q_{\phi}$ preserves the contextual structures between $\bm{x}_0$ and $ \hat{\bm{x}}_{1}^0 $, we introduce additional context-preserving regularization into our training objective besides network design. Various works has shown that the perceptual-level constraints will lead better results~\cite{zhang2018unreasonable,johnson2016perceptual}. Inspired by these works, We chose to use multi-scale structural similarity measure (SSIM)~\cite{brunet2011mathematical} and 
PatchNCE~\cite{park2020contrastive}, arguing that SSIM constraint is defined as:
\begin{equation}
    \begin{split}
        \mathbb{L}_{SSIM}(\phi,t_i) := (\mathbb{L}^{gen}(\bm{x}_{t_i},\hat{\bm{x}}_{1}^{t_i})+\mathbb{L}^{idt}(\bm{x}_1,\hat{\bm{x}}_{1}^1))/2 \\
        \text{s.t}\begin{cases}
            \mathbb{L}^{gen} := 1 - SSIM(\bm{x}_{t_i}, \hat{\bm{x}}_{1}^{t_i}) \\
            \mathbb{L}^{idt} := 1 - SSIM(\bm{x}_1, \hat{\bm{x}}_{1}^1) \\
        \end{cases}
    \end{split}
\end{equation}
where $\mathbb{L}^{gen}$ enforces the consistency of generation and $\mathbb{L}^{idt}$ ensures the generator does not introduce unnecessary changes. The PatchNCE constraint is applied in the same manner as in~\cite{park2020contrastive}. Since SSIM focuses on global perceptual differences, and PatchNCE enforces patch-wise alignment, we argue that they regularize the generation process at different level. Thus, the final objective for time $t_i$ is:
\begin{equation}
\begin{split}
    \mathbb{L}_{CUNSB}(\phi,t_i) :=  &\mathbb{L}_{Adv}(\phi,t_i) + \lambda_{SB} \mathbb{L}_{SB}(\phi,t_i)  \\
     + &\lambda_{S}\mathbb{L}_{SSIM}(\phi,t_i) \\
    + &\lambda_{P}\mathbb{L}_{PatchNCE}(\phi,t_i) \label{eq:Final UNSB}
\end{split}
\end{equation}
where $\lambda_{SB},\lambda_{S}$ and $\lambda_{P}$ are balancing parameters.

\section{Experiments and results}
\label{sec:Experiment}
\subsection{Experimental Details}
\noindent \textbf{CUNSB-RFIE settings}. Our model consists of three types of trainable networks: Markovian discriminators~\cite{isola2017image}, a customized generator and multi-layer perceptrons (MLPs). Specifically, two discriminators estimates $\mathbb{L}_{Adv}$ and $H(q_{\phi_i}(\bm{x}_{t_i}, \bm{x}_1))$ in Eq.~\ref{eq:optim} using adversarial learning strategies and the ideals from~\cite{belghazi2018mutual}, respectively. Nine MLPs, each with two layers, are used to calculate the PatchNCE regularization, as described in~\cite{park2020contrastive}. For other parameters, we set the number of time steps $N=5$, $\lambda_{SB}=1, \lambda_{P}=1, \lambda_{S} = 0.8, \tau=0.01$. 

\noindent \textbf{Evaluation}. We evaluate the capability of our method for image enhancement on three publicly available datasets: EyeQ~\cite{eyeq}, DRIVE~\cite{drive} and IDRID~\cite{idrid}. All images are center-cropped and resized to a resolution of 256 × 256 before training. We use Peak Signal-to-Noise Ratio (PSNR) and Structural Similarity Index Measure (SSIM) to access enhancement quality, and we utilize Area under ROC (AUC), Precision-Recall (PR), Sensitivity (SE), Specificity (SP), and  Jaccard index to evaluate two downstream segmentation tasks.
\begin{table*}[!t]
\centering
\caption{Performance comparison with baselines. The best performance within each column is highlighted in bold.}
\tiny
\resizebox{0.9\textwidth}{!}{%
\begin{tabular}{cccccccc}
\toprule
\multirow{2}{*}{} & \multirow{2}{*}{\textbf{Method}} & \multicolumn{2}{c}{\textbf{EyeQ}} & \multicolumn{2}{c}{\textbf{IDRID}} & \multicolumn{2}{c}{\textbf{DIRVE}} \\ \cmidrule(l){3-8} 
                                          &                                  & \textbf{SSIM} $\uparrow$   & \textbf{PSNR}$\uparrow$   & \textbf{SSIM}    & \textbf{PSNR}   & \textbf{SSIM}    & \textbf{PSNR}   \\ \midrule
\multirow{1}{*}{\textit{Supervised Methods}}       & cofe-Net~\cite{shen2020modeling}                         & 0.880           & 17.25           & 0.65            & 19.07            & 0.66            & 19.11            \\
                                          % & PCE-Net~\cite{10.1007/978-3-031-16434-7_49}                          & 18.54           & 0.874           & 28.94            & 0.646           & 20.26            & 0.775           \\ 
                                          \midrule
\multirow{8}{*}{\textit{Unsupervised Methods}}&I-SECRET~\cite{i-secret} & 0.884 & 14.84 & 0.756 & 18.40 &0.669 &18.75 \\
& GFE-Net~\cite{li2023generic}                          & 0.80           & 18.68            & 0.631            & 19.21  & \textbf{0.715}            & 15.14           \\
                                          & SCR-Net~\cite{scrnet}                          & 0.796           & 18.86           & 0.616            & 19.51           & 0.668            & 18.50           \\
                                          & CycleGAN~\cite{cyclegan}                         & 0.878           & 22.93           & 0.764            & 21.01           & 0.653            & 21.59  \\
                                            % \clines[2-6]
                                            %\cmidrule(l){2-8}
                                          & OTTGAN~\cite{9763342}                           & 0.895           & 23.25           & 0.755            & 20.94           & 0.637            & 20.73           \\
                                          & OTEGAN~\cite{zhu2023otre}                           & \textbf{0.898}           & 23.51           & 0.687            & 18.20           & 0.601            & 17.96           \\ \cmidrule(l){2-8}
                                          & Ours                       & 0.858 & \textbf{27.61} & \textbf{0.768}  & \textbf{21.92}          & 0.649  & \textbf{22.07}          \\ \bottomrule
\end{tabular}%
}
\label{deg-exp}
\end{table*}

\begin{table*}[!t]
    \centering
    \caption{Performance comparison with baselines on blood vessel and diabetic lesions (EX and HE) segmentation tasks on the DRIVE~\cite{drive} and IDRID~\cite{idrid} dataset, respectively.}
    \tiny
    \resizebox{0.9\textwidth}{!}{%
    \begin{tabular}{lcccc|ccc|ccc}
        \toprule
         \multirow{2}[3]{*}{} & \multicolumn{4}{c}{Vessel Segmentation} & \multicolumn{3}{c}{EX} & \multicolumn{3}{c}{HE} \\ 
         \cmidrule(lr){2-5}  \cmidrule(lr){6-8}  \cmidrule(lr){9-11}
          
         Method & ROC $\uparrow$ & PR $\uparrow$ & SE $\uparrow$ & SP $\uparrow$ & ROC  & PR & Jaccard $\uparrow$ & ROC & PR & Jaccard\\ \midrule
        cofe-Net~\cite{shen2020modeling} & 0.911 & 0.766 & \textbf{0.624} & 0.977 & 0.898 & 0.257 & 0.190 & 0.807 & 0.083 & 0.076\\
        I-SECRET~\cite{i-secret} & 0.878 & 0.695 & 0.531 & 0.977 & 0.900 & 0.257 & 0.193 & 0.821 & 0.080 & 0.083 \\
        % PCE-Net~\cite{10.1007/978-3-031-16434-7_49} & 0.879 & 0.696 & 0.540 & 0.976 & 0.949 & 0.500 & 0.328 & 0.874 & 0.102 & 0.134\\
        GFE-Net~\cite{li2023generic} & 0.911 & 0.762 & 0.619 & 0.977 & 0.904 & 0.296 & 0.205 & 0.809 & 0.087 & 0.069\\
        SCR-Net~\cite{scrnet} & 0.904 & 0.748 & 0.599 & 0.977 & 0.906 & 0.259 & 0.181 & 0.830 & 0.118 & 0.107 \\
        CycleGAN~\cite{cyclegan} & 0.885 & 0.718 & 0.580 & 0.975  & 0.906 & 0.379 & 0.179 & 0.809 & 0.123 & 0.101\\
        OTTGAN~\cite{9763342} & 0.896 & 0.740 & 0.592 & 0.977  & 0.928 & 0.438 & 0.239 & 0.872 & 0.197 & \textbf{0.140} \\
        OTEGAN~\cite{zhu2023otre} & 0.908 & 0.764 & 0.623 & 0.977 & \textbf{0.953} & \textbf{0.513} & \textbf{0.338} & \textbf{0.882} & \textbf{0.242} & 0.133 \\
        \midrule
        Ours & \textbf{0.918} & \textbf{0.771} & 0.607 & \textbf{0.979} & 0.921 & 0.327 & 0.215 & 0.856 & 0.122 & 0.115 \\
        \bottomrule
    \end{tabular}}
    \label{tab-seg}
    \vspace{-0.3cm}
\end{table*}
\noindent \textbf{Baselines}. We compare our methods against the following baselines: \textit{Supervised algorithm}: cofe-Net \cite{shen2020modeling}, \textit{Unsupervised or GAN based algorithms}: GFE-Net \cite{li2023generic}, CycleGAN \cite{cyclegan}, SCR-Net \cite{scrnet}, I-SECRET \cite{i-secret}, and \textit{OT based techniques}: OTTGAN \cite{9763342} and OTEGAN \cite{zhu2023otre}. For image enhancement tasks, models are trained on the EyeQ dataset, and the trained weights are directly evaluated on the IDRID and DRIVE datasets. All models are used to perform vessel segmentation tasks on the DRIVE dataset and diabetic lesion segmentation, including Hard Exudates (EX) and Hemorrhages (HE), on the IDRID dataset. More details on the experimental process are provided in the appendix.

\begin{figure*}[h]
\centering
\includegraphics[width=0.8\textwidth]{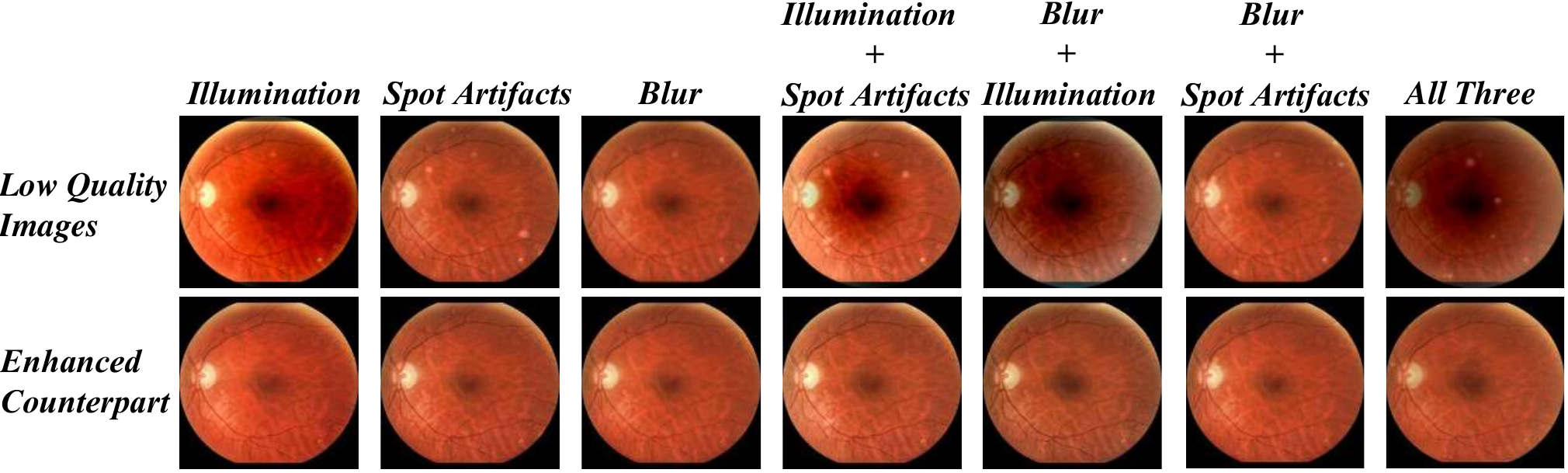}  % Replace with your figure file and remove the example
\vspace{-0.4cm}
\caption{Qualitative illustration of generation results over different noise (i.e., illumination, spot artifacts, and blurring). Our pipeline achieves good performance even on images with mixed noise (\textbf{col.} 4, 5, and 6). }
\label{CUNSB-degrded}
\end{figure*}

\begin{figure*}[!t]
  \centering
  \includegraphics[width=1.0\textwidth]{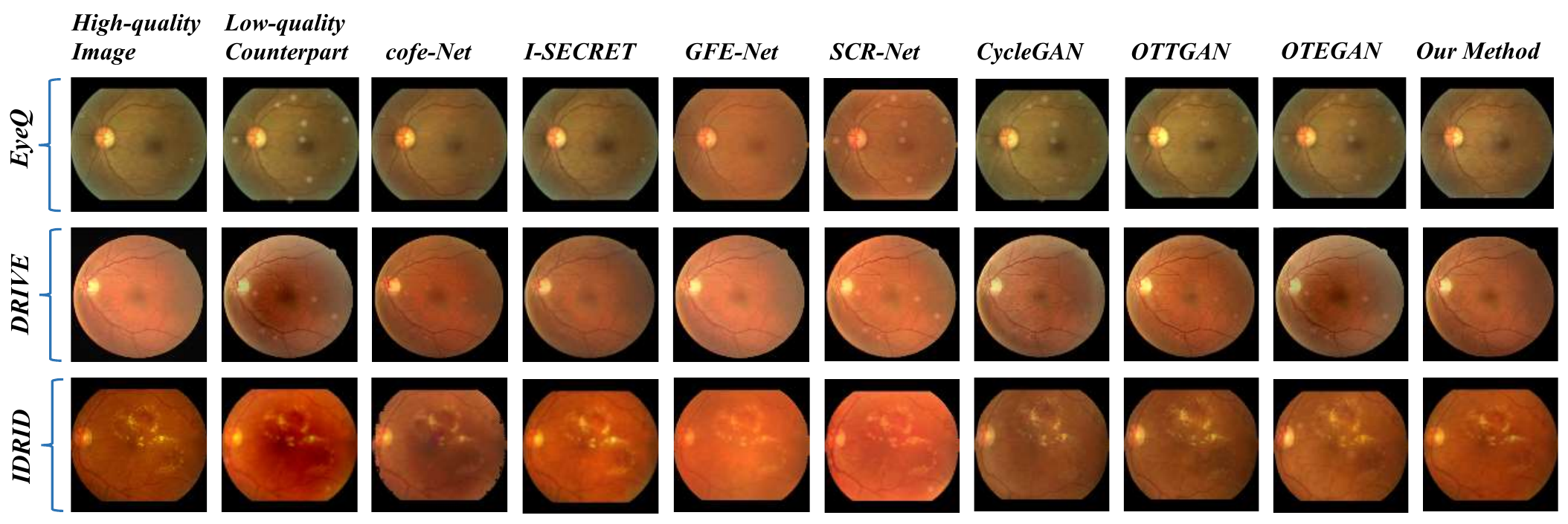}  % Replace with your figure file and remove the example
  \caption{Visual comparison of our method with baselines across three datasets. \textbf{Col 1} represents the high-quality ground truth, \textbf{Col 2} represents the degraded images, \textbf{Col 10} shows our results, and the columns in between show the enhanced results from other models. }
  \label{CUNSB-enhanced-result}
\end{figure*}

\begin{figure*}[!t]
  \centering
  \includegraphics[width=1.0\textwidth]{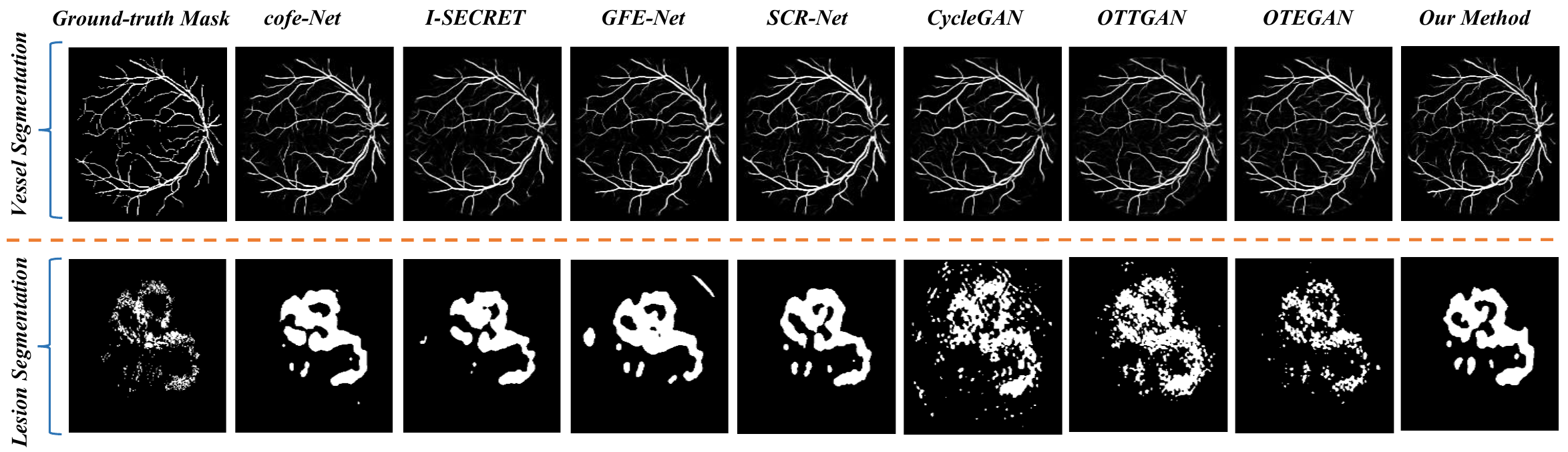}  % Replace with your figure file and remove the example
  \caption{Vessel and Lesion (Hard Exudates) segmentation performance over the enhanced retinal images from different models. \textbf{Col} 1 represents the ground-truth mask, and the left columns represent the generated masks from corresponding models.}
  \label{CUNSB-seg-illustration}
\end{figure*}

\begin{figure*}[!t]
  \centering
  \includegraphics[width=1.0\textwidth]{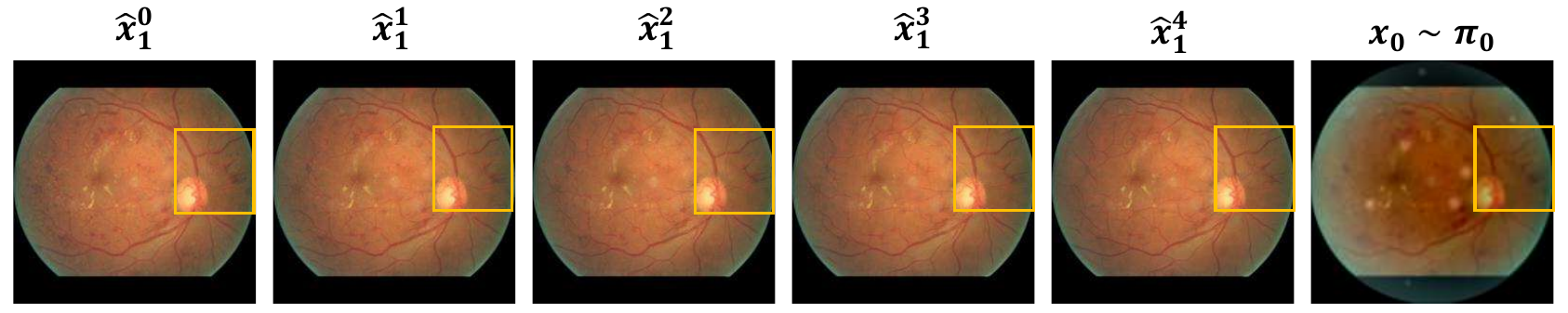}  % Replace with your figure file and remove the example
  \vspace{-0.4cm}
  \caption{ Illustration of smooth influence in CUNSB-RFIE generation. \textbf{Col} 1 to 5 represents synthetic high-quality images at different steps, and 6 represents the low-quality input. The orange box highlights the region that becomes smoothed during the iterative process}
  \label{CUNSB-smooth}
\end{figure*}

\begin{table}[ht]
\centering
% \scriptsize  % Use \scriptsize for smaller font
\resizebox{1.0\columnwidth}{!}
{
%  % Adjust width to better fit the column
\begin{tabular}{ccccc}
\hline
\textbf{DSC} & \textbf{PatchNCE Reg} & \textbf{SSIM Reg} & \textbf{PSNR} & \textbf{SSIM} \\ \hline
$\times$ & $\times$ & $\times$ & 22.232 & 0.793 \\
$\checkmark$ & $\times$ & $\times$ & 25.717 & 0.848 \\
$\checkmark$ & $\times$ & $\checkmark$ & 26.846 & 0.854 \\
$\checkmark$ & $\checkmark$ & $\checkmark$ & 27.611 & 0.858 \\ \hline
\end{tabular}
}
\caption{Ablation Study over DSC, PatchNCE regularization and SSIM regularization}
\label{tab:ablation-study}
\end{table}

\subsection{Experimental Results}
\noindent \textbf{Results in enhancement experiments}.
\noindent We demonstrated the effectiveness of the proposed method across various degradation cases in Fig.~\ref{CUNSB-degrded}. Our model exhibits robust denoising capabilities, even in the most challenging case, which involves the combination of all three types of noise. Additionally, it ensures high generation fidelity by preserving low-level details, such as the background and vessel structure.

\noindent A qualitative comparison and illustration with baselines across three common public datasets is shown in Tab.~\ref{deg-exp} and Fig.~\ref{CUNSB-enhanced-result}, respectively. Specifically, on the EyeQ dataset, our model significantly improves PSNR, suggesting that low-frequency features, such as the background and blood vessels, are well-preserved while maintaining compatible performance over SSIM score with baselines. As illustrated in the first row of Fig.~\ref{CUNSB-enhanced-result}, some baselines (i.e., GFE-Net and SCR-Net) fail to generate realistic images, resulting in noticeable background alterations. Other baselines (i.e., OTTGAN and OTEGAN), although achieving high SSIM score, fail to correctly remove spot artifacts.  

\noindent Additionally, our SB-based model demonstrates stability, image-level fidelity, and strong generalization capability across different image domains, achieving the highest PSNR and SSIM scores on the IDRID dataset and the best PSNR metric on the DRIVE dataset. As shown in the second and third rows in Fig.~\ref{CUNSB-enhanced-result}, some models, particularly OT-based GANs (i.e., OTEGAN and OTTGAN), which show advantages in modeling complex, localized, high-frequency information in retinal images through one-step mapping, struggle with domain shifts and pixel-wise accuracy, leading to the preservation of spot artifacts. Other baselines (i.e., GFE-Net, SCR-Net and I-SECRET) display noticeable inconsistencies. 

\noindent However, we observe a clear gap in SSIM metrics on the EyeQ dataset. We argue that this is because the iterative distribution transportation inevitably smooths out those high-frequency information. Further details will be discussed in Sec.~\ref{sec:discussion}.

\noindent \textbf{Results in segmentation tasks}. Experiments on two segmentation tasks further demonstrate the superiority of CUNSB-RFIE in capturing low-level features. The results are presented in Tab.~\ref{tab-seg}. Specifically, our methods outperformed others in the vessel segmentation task, achieving the best ROC, PR and SP scores. In Fig.~\ref{CUNSB-seg-illustration}, it is clear that most baselines (i.e., OTTGAN, CylceGAN) fails to predict the tiny peripheral vessels, while others, such as OTEGAN, introduce nonexistent connections. 

\noindent In the lesion segmentation task, OTTGAN and OTEGAN show superiority in capturing punctate features. While CycleGAN correctly recognizes lesions, it introduces too many nonexistent noisy points. Our methods face the challenge of over smoothing high-frequency lesion structures. As a result, while the overall morphology is preserved, the synthetic mask tends to be overly continuous.

\subsection{Ablation studies} \noindent We discuss the ablation study we conducted. Unlike the vanilla SB (corresponding to the solution of Eq.~\ref{eq:UNSB}), CUNSB-RFIE leverages DSC to preserve vessel information during training and incorporates task-specific regularization to ensure generation consistency. To assess the impact of these factors on the final generation performance, we performed ablation studies on the EyeQ dataset, with the results reported in Tab.~\ref{tab:ablation-study}. We observed poor results when neither DSC nor any regularization is used in the training pipeline. Next, when we modified the generator architecture by incorporating DSC, the results improved significantly, even surpassing some baselines. This indicates that DSC is beneficial for learning small but important features in retinal images. Finally, we added the PatchNCE and SSIM regularizations to the generator, and the results gradually improved, reaching the best performance with CUNSB-RFIE. This implies that DSC and two regularizations work in orthogonal ways.

\section{Discussion}\label{sec:discussion}
\noindent The trained generator $q_\phi(\bm{x}|\bm{x}_{t_i})$ and Eq.~\ref{eq: sb_dis_any} provide a framework for iteratively refining high-quality images, denoted as $\hat{\bm{x}}_{1}^{t_i}$ for $t_i \in \{ 0,1,2,3,4\}$. In~\cite{kim2023unpaired}, it is proposed to use $\hat{\bm{x}}_{1}^{4}$ as the final generation result for natural images, achieving the best quality. However, we observe a counterintuitive phenomenon for retinal images (denoted in appendix). As the time steps $t_i$ increase, the SSIM and PSNR between the enhanced images and their corresonding ground-truth gradually decrease. We argue that this is due to \textbf{high-frequency information loss in distributions transfer process}. Several works~\cite{meng2021sdedit,choi2021ilvr,si2024freeu} have shown that diffusion models face challenges in generating images with desired semantic information. This is because their iterative generation process, while effective at smoothly modeling complex distribution transitions, tends to introduce more stochasticity due to the accumulation of Gaussian noise. As a result, high-frequency components, which are highly sensitive to noise, are progressively smoothed as time increases. In our pipeline, CUNSB-RFIE, where the training involves iterative sampling at randomly selected time step $t_j$ for every training iteration (shown in Fig.~\ref{CUNSB-structure} (a)), this smoothing effect inevitably leads to the degradation of high-frequency features in retinal images, such as lesion structures (see Fig.~\ref{CUNSB-smooth}). Consequently, this results in an inferior performance in SSIM and lesion segmentation tasks compared to some GAN-based methods (e.g., OTTGAN and OTEGAN). On the other hand, the smooth distribution transition helps maintain pixel-wise alignment, which contributes to higher PSNR values and better performance in vessel segmentation compared to GAN-based methods.

\section{Conclusion}
\noindent In this study, we propose a Context-aware Unpaired Neural Schr\"{o}dinger Bridge (CUNSB-RFIE) pipeline for the enhancement of retinal fundus images. To preserve blood vessel information in retinal images, we utilize Dynamic Snake Convolution, enabling our generator to focus more on low-frequency features.  To further ensure consistency in generation, we introduce context-preserving regularization by combining PatchNCE and SSIM constraints. We demonstrated the effectiveness and generalizability of CUNSB-RFIE in enhancing images while studying vessel structures across three datasets. Although our method shows limitations in preserving lesion structures, it still demonstrates the potential for broader applications in medical image enhancement using SB-based models. In future research, we anticipate further exploration to improve our model's ability to capture localized high-frequency information.

\section{Acknowledgments} 
\noindent This work was supported by grants from NIH (R01EY032125 and R01DE030286), the State of Arizona via the Arizona Alzheimer Consortium.
% \textbf{Acknowledgments.}   The work was partially supported by NIH grants (R21AG065942, R01EY032125, and R01DE030286). 

%%%%%%%%% REFERENCES
{\small
\bibliographystyle{ieee_fullname}
\bibliography{egbib}

\begin{thebibliography}{10}\itemsep=-1pt

\bibitem{belghazi2018mutual}
Mohamed~Ishmael Belghazi, Aristide Baratin, Sai Rajeshwar, Sherjil Ozair, Yoshua Bengio, Aaron Courville, and Devon Hjelm.
\newblock Mutual information neural estimation.
\newblock In {\em International conference on machine learning}, pages 531--540. PMLR, 2018.

\bibitem{brunet2011mathematical}
Dominique Brunet, Edward~R Vrscay, and Zhou Wang.
\newblock On the mathematical properties of the structural similarity index.
\newblock {\em IEEE Transactions on Image Processing}, 21(4):1488--1499, 2011.

\bibitem{i-secret}
Pujin Cheng, Li Lin, Yijin Huang, Junyan Lyu, and Xiaoying Tang.
\newblock I-secret: Importance-guided fundus image enhancement via semi-supervised contrastive constraining.
\newblock In {\em Medical Image Computing and Computer Assisted Intervention--MICCAI 2021: 24th International Conference, Strasbourg, France, September 27--October 1, 2021, Proceedings, Part VIII 24}, pages 87--96. Springer, 2021.

\bibitem{choi2021ilvr}
Jooyoung Choi, Sungwon Kim, Yonghyun Jeong, Youngjune Gwon, and Sungroh Yoon.
\newblock Ilvr: Conditioning method for denoising diffusion probabilistic models.
\newblock {\em arXiv preprint arXiv:2108.02938}, 2021.

\bibitem{dai2017deformable}
Jifeng Dai, Haozhi Qi, Yuwen Xiong, Yi Li, Guodong Zhang, Han Hu, and Yichen Wei.
\newblock Deformable convolutional networks.
\newblock In {\em Proceedings of the IEEE international conference on computer vision}, pages 764--773, 2017.

\bibitem{de2021diffusion}
Valentin De~Bortoli, James Thornton, Jeremy Heng, and Arnaud Doucet.
\newblock Diffusion schr{\"o}dinger bridge with applications to score-based generative modeling.
\newblock {\em Advances in Neural Information Processing Systems}, 34:17695--17709, 2021.

\bibitem{deeplearning5}
Oana~M. Dumitrascu, Xin Li, Wenhui Zhu, Bryan~K. Woodruff, Simona Nikolova, Jacob Sobczak, Amal Youssef, Siddhant Saxena, Janine Andreev, Richard~J. Caselli, John~J. Chen, and Yalin Wang.
\newblock Color fundus photography and deep learning applications in alzheimer’s disease.
\newblock {\em Mayo Clinic Proceedings: Digital Health}, 2024.

\bibitem{deeplearning3}
Oana~M Dumitrascu, Wenhui Zhu, Peijie Qiu, Keshav Nandakumar, and Yalin Wang.
\newblock Automated retinal imaging analysis for alzheimers disease screening.
\newblock In {\em IEEE International Symposium on Biomedical Imaging: From Nano to Macro (ISBI)}, 2022.

\bibitem{eyeq}
Huazhu Fu, Boyang Wang, Jianbing Shen, and et al.
\newblock Evaluation of retinal image quality assessment networks in different color-spaces.
\newblock {\em MICCAI}, pages 48--56, 2019.

\bibitem{goodfellow2014generative}
Ian~J. Goodfellow, Jean Pouget-Abadie, Mehdi Mirza, Bing Xu, David Warde-Farley, Sherjil Ozair, Aaron Courville, and Yoshua Bengio.
\newblock Generative adversarial networks, 2014.

\bibitem{gu2023optimal}
Xiang Gu, Liwei Yang, Jian Sun, and Zongben Xu.
\newblock Optimal transport-guided conditional score-based diffusion model.
\newblock {\em Advances in Neural Information Processing Systems}, 36:36540--36552, 2023.

\bibitem{ho2020denoising}
Jonathan Ho, Ajay Jain, and Pieter Abbeel.
\newblock Denoising diffusion probabilistic models.
\newblock {\em Advances in neural information processing systems}, 33:6840--6851, 2020.

\bibitem{isola2017image}
Phillip Isola, Jun-Yan Zhu, Tinghui Zhou, and Alexei~A Efros.
\newblock Image-to-image translation with conditional adversarial networks.
\newblock In {\em Proceedings of the IEEE conference on computer vision and pattern recognition}, pages 1125--1134, 2017.

\bibitem{johnson2016perceptual}
Justin Johnson, Alexandre Alahi, and Li Fei-Fei.
\newblock Perceptual losses for real-time style transfer and super-resolution.
\newblock In {\em Computer Vision--ECCV 2016: 14th European Conference, Amsterdam, The Netherlands, October 11-14, 2016, Proceedings, Part II 14}, pages 694--711. Springer, 2016.

\bibitem{kim2023unpaired}
Beomsu Kim, Gihyun Kwon, Kwanyoung Kim, and Jong~Chul Ye.
\newblock Unpaired image-to-image translation via neural schr$\backslash$" odinger bridge.
\newblock {\em arXiv preprint arXiv:2305.15086}, 2023.

\bibitem{leonard2013survey}
Christian L{\'e}onard.
\newblock A survey of the schr $\backslash"$ odinger problem and some of its connections with optimal transport.
\newblock {\em arXiv preprint arXiv:1308.0215}, 2013.

\bibitem{scrnet}
Heng Li, Haofeng Liu, Huazhu Fu, Hai Shu, Yitian Zhao, Xiaoling Luo, Yan Hu, and Jiang Liu.
\newblock Structure-consistent restoration network for cataract fundus image enhancement.
\newblock In Linwei Wang, Qi Dou, P.~Thomas Fletcher, Stefanie Speidel, and Shuo Li, editors, {\em Medical Image Computing and Computer Assisted Intervention -- MICCAI 2022}, pages 487--496, Cham, 2022. Springer Nature Switzerland.

\bibitem{li2023generic}
Heng Li, Haofeng Liu, Huazhu Fu, Yanwu Xu, Hai Shu, Ke Niu, Yan Hu, and Jiang Liu.
\newblock A generic fundus image enhancement network boosted by frequency self-supervised representation learning.
\newblock {\em Medical Image Analysis}, 90:102945, 2023.

\bibitem{li2022annotation}
Heng Li, Haofeng Liu, Yan Hu, Huazhu Fu, Yitian Zhao, Hanpei Miao, and Jiang Liu.
\newblock An annotation-free restoration network for cataractous fundus images.
\newblock {\em IEEE Transactions on Medical Imaging}, 41(7):1699--1710, 2022.

\bibitem{li2024diffusion}
Muheng Li, Xia Li, Sairos Safai, Damien Weber, Antony Lomax, and Ye Zhang.
\newblock Diffusion schr$\backslash$" odinger bridge models for high-quality mr-to-ct synthesis for head and neck proton treatment planning.
\newblock {\em arXiv preprint arXiv:2404.11741}, 2024.

\bibitem{liu2023esdiff}
Fengting Liu and Wenhui Huang.
\newblock Esdiff: a joint model for low-quality retinal image enhancement and vessel segmentation using a diffusion model.
\newblock {\em Biomedical Optics Express}, 14(12):6563--6578, 2023.

\bibitem{liu20232}
Guan-Horng Liu, Arash Vahdat, De-An Huang, Evangelos~A Theodorou, Weili Nie, and Anima Anandkumar.
\newblock I$^{2}$sb: Image-to-image schr{\"o}dinger bridge.
\newblock {\em arXiv preprint arXiv:2302.05872}, 2023.

\bibitem{10.1007/978-3-031-16434-7_49}
Haofeng Liu, Heng Li, Huazhu Fu, Ruoxiu Xiao, Yunshu Gao, Yan Hu, and Jiang Liu.
\newblock Degradation-invariant enhancement of fundus images via pyramid constraint network.
\newblock In Linwei Wang, Qi Dou, P.~Thomas Fletcher, Stefanie Speidel, and Shuo Li, editors, {\em Medical Image Computing and Computer Assisted Intervention -- MICCAI 2022}, pages 507--516, Cham, 2022. Springer Nature Switzerland.

\bibitem{meng2021sdedit}
Chenlin Meng, Yutong He, Yang Song, Jiaming Song, Jiajun Wu, Jun-Yan Zhu, and Stefano Ermon.
\newblock Sdedit: Guided image synthesis and editing with stochastic differential equations.
\newblock {\em arXiv preprint arXiv:2108.01073}, 2021.

\bibitem{park2020contrastive}
Taesung Park, Alexei~A Efros, Richard Zhang, and Jun-Yan Zhu.
\newblock Contrastive learning for unpaired image-to-image translation.
\newblock In {\em Computer Vision--ECCV 2020: 16th European Conference, Glasgow, UK, August 23--28, 2020, Proceedings, Part IX 16}, pages 319--345. Springer, 2020.

\bibitem{41493}
E. Peli and T. Peli.
\newblock Restoration of retinal images obtained through cataracts.
\newblock {\em IEEE Transactions on Medical Imaging}, 8(4):401--406, 1989.

\bibitem{idrid}
Prasanna Porwal and et al.
\newblock Idrid: A database for diabetic retinopathy screening research.
\newblock {\em Data}, 3(3), 2018.

\bibitem{qi2023dynamic}
Yaolei Qi, Yuting He, Xiaoming Qi, Yuan Zhang, and Guanyu Yang.
\newblock Dynamic snake convolution based on topological geometric constraints for tubular structure segmentation.
\newblock In {\em Proceedings of the IEEE/CVF International Conference on Computer Vision}, pages 6070--6079, 2023.

\bibitem{rombach2022high}
Robin Rombach, Andreas Blattmann, Dominik Lorenz, Patrick Esser, and Bj{\"o}rn Ommer.
\newblock High-resolution image synthesis with latent diffusion models.
\newblock In {\em Proceedings of the IEEE/CVF conference on computer vision and pattern recognition}, pages 10684--10695, 2022.

\bibitem{salmona2022can}
Antoine Salmona, Valentin De~Bortoli, Julie Delon, and Agnes Desolneux.
\newblock Can push-forward generative models fit multimodal distributions?
\newblock {\em Advances in Neural Information Processing Systems}, 35:10766--10779, 2022.

\bibitem{schrodinger1932theorie}
Erwin Schr{\"o}dinger.
\newblock Sur la th{\'e}orie relativiste de l'{\'e}lectron et l'interpr{\'e}tation de la m{\'e}canique quantique.
\newblock {\em Annales de l'institut Henri Poincar{\'e}}, 2(4):269--310, 1932.

\bibitem{selim2023latent}
Md Selim, Jie Zhang, Faraneh Fathi, Michael~A Brooks, Ge Wang, Guoqiang Yu, and Jin Chen.
\newblock Latent diffusion model for medical image standardization and enhancement.
\newblock {\em arXiv preprint arXiv:2310.05237}, 2023.

\bibitem{shao2024prior}
Zhuchen Shao, Mark~A Anastasio, and Hua Li.
\newblock Prior-guided diffusion model for cell segmentation in quantitative phase imaging.
\newblock {\em arXiv preprint arXiv:2405.06175}, 2024.

\bibitem{shen2020modeling}
Z. Shen, H. Fu, J. Shen, and L. Shao.
\newblock {{M}odeling and {E}nhancing {L}ow-{Q}uality {R}etinal {F}undus {I}mages}.
\newblock {\em IEEE Trans Med Imaging}, 40(3):996--1006, 2021.

\bibitem{shi2024diffusion}
Yuyang Shi, Valentin De~Bortoli, Andrew Campbell, and Arnaud Doucet.
\newblock Diffusion schr{\"o}dinger bridge matching.
\newblock {\em Advances in Neural Information Processing Systems}, 36, 2024.

\bibitem{si2024freeu}
Chenyang Si, Ziqi Huang, Yuming Jiang, and Ziwei Liu.
\newblock Freeu: Free lunch in diffusion u-net.
\newblock In {\em Proceedings of the IEEE/CVF Conference on Computer Vision and Pattern Recognition}, pages 4733--4743, 2024.

\bibitem{sohl2015deep}
Jascha Sohl-Dickstein, Eric Weiss, Niru Maheswaranathan, and Surya Ganguli.
\newblock Deep unsupervised learning using nonequilibrium thermodynamics.
\newblock In {\em International conference on machine learning}, pages 2256--2265. PMLR, 2015.

\bibitem{song2020denoising}
Jiaming Song, Chenlin Meng, and Stefano Ermon.
\newblock Denoising diffusion implicit models.
\newblock {\em arXiv preprint arXiv:2010.02502}, 2020.

\bibitem{drive}
J. Staal and et al.
\newblock {{R}idge-based vessel segmentation in color images of the retina}.
\newblock {\em IEEE Trans Med Imaging}, 23(4):501--509, 2004.

\bibitem{tong2023improving}
Alexander Tong, Nikolay Malkin, Guillaume Huguet, Yanlei Zhang, Jarrid Rector-Brooks, Kilian Fatras, Guy Wolf, and Yoshua Bengio.
\newblock Improving and generalizing flow-based generative models with minibatch optimal transport.
\newblock {\em arXiv preprint arXiv:2302.00482}, 2023.

\bibitem{vargas2021solving}
Francisco Vargas, Pierre Thodoroff, Austen Lamacraft, and Neil Lawrence.
\newblock Solving schr{\"o}dinger bridges via maximum likelihood.
\newblock {\em Entropy}, 23(9):1134, 2021.

\bibitem{9763342}
Wei Wang, Fei Wen, Zeyu Yan, and Peilin Liu.
\newblock Optimal transport for unsupervised denoising learning.
\newblock {\em IEEE PAMI}, pages 1--1, 2022.

\bibitem{wang2024implicit}
Yuang Wang, Siyeop Yoon, Pengfei Jin, Matthew Tivnan, Zhennong Chen, Rui Hu, Li Zhang, Zhiqiang Chen, Quanzheng Li, and Dufan Wu.
\newblock Implicit image-to-image schrodinger bridge for ct super-resolution and denoising.
\newblock {\em arXiv preprint arXiv:2403.06069}, 2024.

\bibitem{wu2023latent}
Chen~Henry Wu and Fernando De~la Torre.
\newblock A latent space of stochastic diffusion models for zero-shot image editing and guidance.
\newblock In {\em Proceedings of the IEEE/CVF International Conference on Computer Vision}, pages 7378--7387, 2023.

\bibitem{xiao2021tackling}
Zhisheng Xiao, Karsten Kreis, and Arash Vahdat.
\newblock Tackling the generative learning trilemma with denoising diffusion gans.
\newblock {\em arXiv preprint arXiv:2112.07804}, 2021.

\bibitem{yu2017dilated}
Fisher Yu, Vladlen Koltun, and Thomas Funkhouser.
\newblock Dilated residual networks.
\newblock In {\em Proceedings of the IEEE conference on computer vision and pattern recognition}, pages 472--480, 2017.

\bibitem{zhang2018unreasonable}
Richard Zhang, Phillip Isola, Alexei~A Efros, Eli Shechtman, and Oliver Wang.
\newblock The unreasonable effectiveness of deep features as a perceptual metric.
\newblock In {\em Proceedings of the IEEE conference on computer vision and pattern recognition}, pages 586--595, 2018.

\bibitem{cyclegan}
Jun{-}Yan Zhu, Taesung Park, Phillip Isola, and Alexei~A. Efros.
\newblock {Unpaired Image-to-Image Translation Using Cycle-Consistent Adversarial Networks}.
\newblock {\em {CVPR}}, pages 2242--2251, 2017.

\bibitem{deeplearning4}
Wenhui Zhu, Peijie Qiu, Xiwen Chen, Huayu Li, Hao Wang, Natasha Lepore, Oana~M Dumitrascu, and Yalin Wang.
\newblock Beyond mobilenet: An improved mobilenet for retinal diseases.
\newblock In {\em International Conference on Medical Image Computing and Computer-Assisted Intervention}, pages 56--65. Springer, 2023.

\bibitem{deeplearning1}
Wenhui Zhu, Peijie Qiu, Xiwen Chen, Xin Li, Natasha Lepore, Oana~M. Dumitrascu, and Yalin Wang.
\newblock nnmobilenet: Rethinking cnn for retinopathy research.
\newblock In {\em Proceedings of the IEEE/CVF Conference on Computer Vision and Pattern Recognition (CVPR) Workshops}, pages 2285--2294, June 2024.

\bibitem{zhu2023otre}
Wenhui Zhu, Peijie Qiu, Oana~M Dumitrascu, Jacob~M Sobczak, Mohammad Farazi, Zhangsihao Yang, Keshav Nandakumar, and Yalin Wang.
\newblock Otre: Where optimal transport guided unpaired image-to-image translation meets regularization by enhancing.
\newblock In {\em International Conference on Information Processing in Medical Imaging}, pages 415--427. Springer, 2023.

\bibitem{zhu2023optimal}
Wenhui Zhu, Peijie Qiu, Mohammad Farazi, Keshav Nandakumar, Oana~M. Dumitrascu, and Yalin Wang.
\newblock Optimal transport guided unsupervised learning for enhancing low-quality retinal images, 2023.

\bibitem{deeplearning2}
Wenhui Zhu, Peijie Qiu, Natasha Lepore, Oana~M Dumitrascu, and Yalin Wang.
\newblock Self-supervised equivariant regularization reconciles multiple-instance learning: Joint referable diabetic retinopathy classification and lesion segmentation.
\newblock In {\em 18th International Symposium on Medical Information Processing and Analysis}, volume 12567, pages 100--107. SPIE, 2023.

\end{thebibliography}
}

\end{document}

%%%%%%%%% TITLE - PLEASE UPDATE
\title{Supplementary for CUNSB-RFIE: Context-aware Unpaired Neural Schr\"{o}dinger Bridge in Retinal Fundus Image Enhancement}

\maketitle

%%%%%%%%% BODY TEXT
\appendix
\section{Experiment Details}
\noindent \textbf{Image Enhancement Experiment}. The experiment was conducted on the synthetically degraded retinal fundus images, created by combining Light Transmission Disturbance, Image Blurring, and Retinal Artifact. The training and testing sets consisted of 3500 and 1891 high-quality retinal images from the EyeQ dataset, respectively. The model was trained for 130 epochs using the Adam optimizer, with an initial learning rate of \(2 \times 10^{-4}\) and $\beta$ values set to 0.5 and 0.999, respectively. The learning rate was linearly decayed to 0 after running the first 80 epochs and the batch size was set to 8. The trained weights with best performance were then evaluated on complete DRIVE and IDRID datasets. SSIM and PSNR scores were calculated between the synthetically generated high-quality images and the corresponding ground-truth, where the low-quality counterparts were generated using the degrading algorithms described in ~\cite{shen2020modeling}.

\noindent \textbf{Downstream Segmentation}. Two Downstream segmentation tasks were conducted to demonstrate the ability to preserve intricate details in low-quality fundus images after enhancement, focusing on both high-frequency structure (i.e., lesion structure) and low-frequency information (i.e., blood vessel structure). Specifically, a vessel segmentation task was conducted on the DRIVE dataset and a diabetic lesion segmentation was performed on the IDRID dataset. 

\noindent For vessel segmentation, we used the official split of the DRIVE dataset, dividing the training and testing sets equally. Performance was evaluated using the Area under ROC (AUC), Precision-Recall curve (PR), Sensitivity (SE), and Specificity (SP). For lesion segmentation, 54 subjects were used in the training set and 27 subjects in the testing set, with a focus on large lesion blocks such as Hard Exudates (EX) and Hemorrhages (HE). The evaluation metrics for this task were AUC, PR and the Jaccard index. 

\noindent A vanilla UNet~\cite{ronneberger2015unet} model was trained from scratch for both segmentation tasks, respectively. For vessel segmentation, we used the Adam optimizer with cross-entropy loss as the objective function. The initial learning rate and batch size were set to \(5 \times 10^{-5}\) and 64, respectively. For lesion segmentation, the same optimizer was used. And the learning rate was set to \(2 \times 10^{-4}\) with a weight decay equal to \(5 \times 10^{-4}\). The batch size was set to 8 the model was trained for 300 epochs.

\section{Experiment Results}
\begin{figure}[htbp]
  \centering
  \includegraphics[width=1.0\columnwidth]{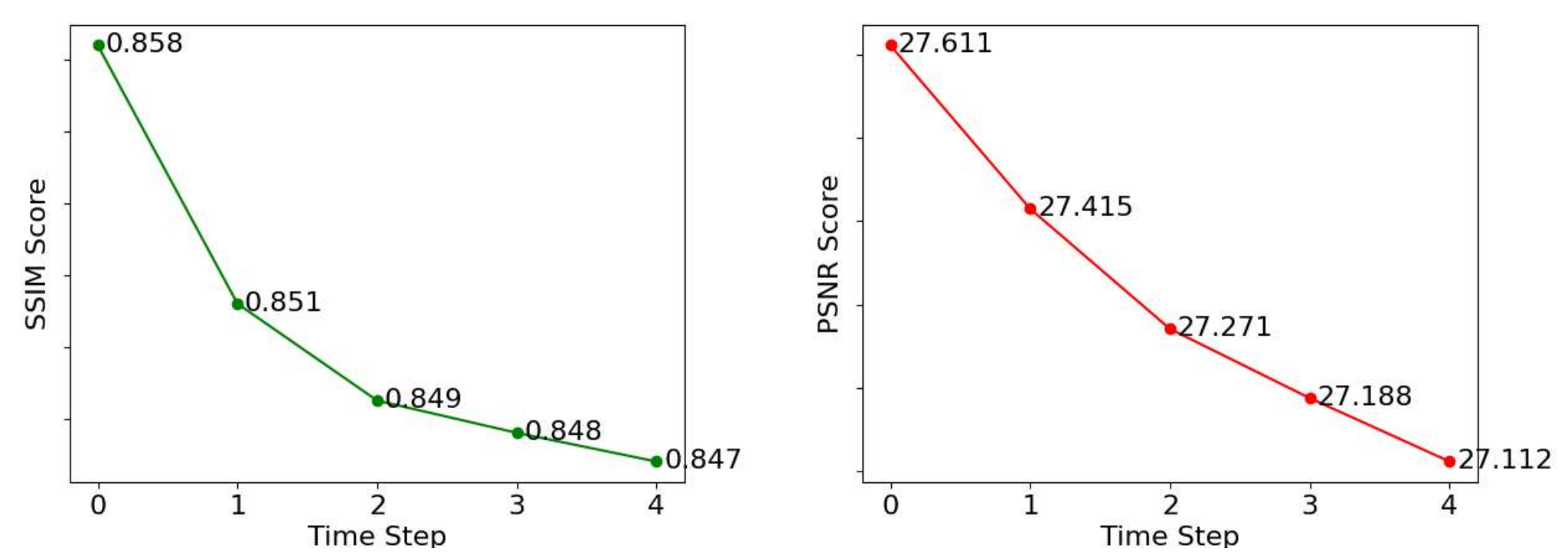}
  \caption{\textbf{Left}: SSIM scores decrease from 0.858 to 0.847 as time increases. \textbf{Right}: PSNR scores decrease from 27.611 to 27.122 as time increases.}
\label{CUNSB-ssim-psnr}
\end{figure}
\noindent From Fig.~\ref{CUNSB-ssim-psnr}, we can find that the quality of synthetic images gradually drop as time step $t_i$ increase, which indicate the smooth influnece shown in CUNSB-RFIE.

%%%%

% \textbf{Acknowledgments.}   The work was partially supported by NIH grants (R21AG065942, R01EY032125, and R01DE030286). 

%%%%%%%%% REFERENCES
{\small
\bibliographystyle{ieee_fullname}
\bibliography{egbib}
}